\documentstyle[eqsecnum,preprint,aps]{revtex}
\begin{document}

\draft

\title{Effects of  distance dependence of exciton hopping on the Davydov
soliton}

\author{E.A. Bartnik}
\address{Faculty of Physics, University of Warsaw,
Hoza 69 Street, 00-681 Warsaw, Poland}

\author{J.A. Tuszy\'nski\thanks{Permanent address: Department of Physics,
University
  of Alberta, Edmonton, Alberta, T6G 2J1, Canada} and D. Sept\footnotemark[1]}
\address{Institut f\"ur Theoretische Physik I,
Heinrich-Heine-Universit\"at D\"usseldorf,\\
D-40225 D\"usseldorf, Germany}

\maketitle

\begin{abstract}

 The Davydov model of energy transfer in molecular chains is reconsidered
assuming the distance dependence of the exciton hopping term.  New equations of
motion for phonons and excitons are derived  within the coherent state
approximation.  Solving these nonlinear equations result in the existence of
Davydov-like solitons.  In the case of a dilatational soliton,   the amplitude
and width is decreased as a results of the mechanism introduced here and above
a critical coupling strength our equations do not allow for localized
solutions.  For compressional solitons,  stability is increased.
\end{abstract}


\section{Introduction}

Davydov\cite{davydov} formulated a model intended to explain an almost lossless
energy transfer in quasi-one-dimensional biomolecular chains such as peptides,
proteins and the DNA.  The key element in the Davydov model is the coupling
between excitons and phonons of the biomolecular chain.  The starting point is
typically the Hamiltonian given by
\begin{equation}
H = \frac{1}{2} \sum_n^{}\left[ \frac{p_n^2}{M} + W (u_n - u_{n-1})^2 \right] +
\sum_{n}^{}
\left[ D A_n^{\dag} A_n - J (A_{n+1}^{\dag} A_n + A_{n+1} A_n^{\dag}) \right]
  \label{hamiltonian}
\end{equation}
where the position of the $n$-th unit is
\begin{equation}
  x_n = d n + u_n
\end{equation}
with $d$ denoting the equilibrium separation and $u_n$ a displacement from
equilibrium.  Here, $p_n$ is the associated momentum, $W$ the elastic modulus
and $M$ the mass of the oscillating unit.  The operators $A_n^{\dag}$ and $A_n$
create and annihilate an exciton at site $n$, respectively.

The constant $D$ describes the energy of a single exciton on a given site while
$J$ refers to hopping of excitons between neighboring sites.  Since both
constants originate from the dipolar nature of the chain's units it is obvious
that they must be distance dependent. Hence, denoting the relative distances
between two neighboring groups as
\begin{equation}
  r_{n,n+1} = |x_{n+1}-x_n| = d + u_{n+1} - u_n
\end{equation}
and
\begin{equation}
  r_{n,n-1} = |x_n-x_{n-1}| = d + u_n - u_{n-1}
\end{equation}
we assume that
\begin{equation}
  D = \Omega + D(r_{n,n+1}) + D(r_{n,n-1})
\end{equation}
and analogously that
\begin{equation}
  J = J(r_{n,n+1}) + J(r_{n,n-1}).
\end{equation}
In most of the literature on the subject only the dependence of $D$ on the
distance has been explicitly introduced.  It is quite obvious, however, that
$J$ should be inversely proportional to the third power of the distance
(dipole-dipole interaction) and this for small displacements gives an
approximately linear dependence of $J$ on the magnitude of the displacement.
Hence, we can postulate that:
\begin{eqnarray}
  D(r_{n,n+1}) & \cong& D_0 + \chi_1 (u_{n+1} - u_n) \\
  D(r_{n,n-1}) & \cong& D_0 + \chi_1 (u_n - u_{n-1}) \\
  J(r_{n,n+1}) & \cong& J_0 + \chi_2 (u_{n+1} - u_n) \\
  J(r_{n,n-1}) & \cong& J_0 + \chi_2 (u_n - u_{n-1})
\end{eqnarray}
For a treatment of similar terms in a somewhat different spirit we refer the
reader to Ref.\cite{zolo}.

First of all, let us note that $\chi_1$ is responsible for the existence of a
localized solution called a Davydov soliton, and this fact is irrespective of
the sign of $\chi_1$ \cite{davydov}.  For $\chi_1>0$, the molecular chain is
locally compressed (compressional soliton) and for $\chi_1<0$, the soliton is
of dilatational type (locally dilatated).  This is not an uncommon situation in
soliton bearing nonlinear lattices \cite{peyrard}.  It is still an open
question as to the value and the sign of $\chi_1$ (see for instance Ref.
\cite{scott2}).   The constant $\chi_2$ is clearly negative since the
interaction diminishes with distance. Scott\cite{scott} had estimated for
peptide chains $\chi_2$ from the approximate relation $\chi_2 \cong3 J/d$ using
$J \cong7.8$ cm$^{-1}$ and $d \cong4.5$ {\AA} as $\chi_2 \equiv 1 pN$.  He also
gave an estimate of $\chi_1 \cong34 pN$.    These values will be used in
subsequent numerical calculations.  We will show in the sections that follow
that keeping both $\chi_1$ and $\chi_2$ in the Hamiltonian results in equations
of motion which differ from the standard Davydov system.  It will be of
interest to examine their solutions and find out whether solitons can be formed
and what affects their properties.

\section{The model Hamiltonian}

Based on the results of the previous section we rewrite the Davydov
Hamiltonian's exciton part as
\begin{eqnarray}
  H_{exc} &=& \sum_n^{} \left[ (\Omega + 2 D_0) + \chi_1(u_{n+1} - u_{n-1})
\right]
A_n^{\dag} A_n \\ \nonumber &-& \sum_n^{} A_n^{\dag} \left[(J_0 +
\chi_2(u_n-u_{n-1}))A_{n-1} +
(J_0 + \chi_2(u_{n+1}-u_n))A_{n+1} \right]
\end{eqnarray}
where the first term is responsible for on-site excitations and the second for
hopping.

In the next stage we find the continuum limit of both the excitonic part
$H_{exc}$ given above and the phonon contribution $H_{ph}$.  To make the
continuum limit we take:
\begin{eqnarray}
  A_n \rightarrow A(x); \,\, A_{n\pm1} \cong A(x) \pm \frac{\partial
A}{\partial x}\; d + \frac{1}{2}
\frac{\partial^2 A}{\partial x^2} \;d^2+ \ldots \\
  u_n \rightarrow u(x); \,\, u_{n\pm1} \cong u(x) \pm \frac{\partial
u}{\partial x} \; d + \frac{1}{2}
\frac{\partial^2 u}{\partial x^2} \; d^2+ \ldots
\end{eqnarray}
and define
\begin{eqnarray}
  J(x) &\equiv& J_0 + \chi_2 \frac{\partial u}{\partial x} \; d\\
D(x) &\equiv& D_0 + \chi_1 \frac{\partial u}{\partial x}\; d.
\end{eqnarray}
As a result, the phonon part of the Hamiltonian becomes:
\begin{equation}
  H_{ph} = \frac{1}{2} \int dx \, \left[ \frac{p^2(x)}{M} + W \left(
\frac{\partial u}{\partial x}
\right)^2\, d^2 \right].
\end{equation}
The exciton contribution to the Davydov Hamiltonian is found as
\begin{eqnarray}
  H_{exc} &=& \int dx \, \left[ \tilde{\Omega} A^{\dag}(x)A(x) + 2 \chi_1 d
\left( \frac{\partial u}%
{\partial x}\right) \; A^{\dag}(x)A(x) \right] \\ \nonumber &-& \int dx \,
\left( J_0 + \chi_2 d%
\left( \frac{\partial u}{\partial x}\right) \right) \left[2 A^{\dag}(x)A(x) -
\frac{\partial
A^{\dag}}{\partial x} \frac{\partial A}{\partial x} \; d^2\right]
\end{eqnarray}
where we have denoted
\begin{equation}
  \tilde{\Omega}  \equiv  \Omega + 2 D_0
\end{equation}
and the hopping part has been obtained from a symmetric form given by
\begin{equation}
  H_{hop} = - \sum_n^{} J_0 (A_n^{\dag} A_{n+1} + A_{n+1}^{\dag} A_n )
\end{equation}
which is explicitly hermitian.

Our next task is to derive equations of motion for the coupled phonon field
$u(x)$ and the exciton field $A(x)$.  This can be done using several
approaches, e.g. through functional minimization, coherent state Ansatz, etc.
In the following section we derive these equations by treating phonons
classically and taking a special Ansatz for excitons.

\section{Deriving the equations of motion}

In order to derive equations of motion we assume the phonons to be virtually
classical so that Hamilton's equations can be applied to $u(x)$ and $p(x)$.
The exciton part is approximated by the expectation value of $H_{exc}$ in the
trial ground state
\begin{equation}
  |\psi\rangle = \int  dx \, a(x) A^{\dag}(x) |0\rangle
\end{equation}
where $a(x)$ is a complex probability amplitude of an exciton wave and
$|0\rangle$is the vacuum.  Thus, the effective Hamiltonian used is
\begin{equation}
  H_{eff} = H_{ph} + \langle \psi|H_{exc}|\psi\rangle
\end{equation}
or explicitly:
\begin{eqnarray}
  H_{eff} & = & \frac{1}{2} \int dx \, \left[\frac{p^2(x)}{M} + W
  \left( \frac{\partial u}{\partial x} \right)^2 \, d^2\right] \\ \nonumber
  &+& \int dx \, \left\{ \left[ (-2 J_0 + \tilde{\Omega}) +
  2(\chi_1-\chi_2) \, d\frac{\partial u}{\partial x} \right] |a(x)|^2 +
  (J_0 +\chi_2 \; d\frac{\partial u}{\partial x} ) \left| \frac{\partial
    a}{\partial x} \right|^2 \, d^2\right\}. \label{ham-eff}
\end{eqnarray}
We now derive the equations of motion for the phonon field as
\begin{equation}
  \frac{\partial u}{\partial t} = \{ H_{eff},p \} = \frac{p(x)}{M}
\label{phononu}
\end{equation}
and
\begin{equation}
  \frac{\partial p}{\partial t} = \{H_{eff},u\} = W d^2  \frac{\partial^2
    u}{\partial x^2} + \frac{\partial}{\partial x} \left[
  2(\chi_1-\chi_2) \, d |a|^2 + \chi_2 \, d^3\left|\frac{\partial a} {\partial
x}
\right|^2 \right]
\label{phononp}
\end{equation}
where the curly brackets above indicate Poisson commutators. Differentiating
eq. (\ref{phononu}) with respect to time and utilizing
eq. (\ref{phononp}) we obtain
\begin{equation}
  \left[ \frac{\partial^2}{\partial t^2} - v_0^2
  \frac{\partial^2}{\partial x^2} \right] u(x,t) = \frac{1}{M}
  \frac{\partial}{\partial x} \left[ 2(\chi_1-\chi_2)\, d |a|^2 + \chi_2 \, d^3
  \left|\frac{\partial a} {\partial x} \right|^2 \right]
  \label{wave-eq}
\end{equation}
which is a wave equation for the displacement field with an additional term
arising due to the coupling with excitons.  The sound velocity $v_0$ is defined
as $v_0 = d \sqrt{W/M}$.

We treat excitons quantum mechanically and hence use the Schr\"odinger equation
\begin{equation}
  i \hbar \frac{\partial}{\partial t} |\psi\rangle = H_{exc} |\psi\rangle
\end{equation}
as an equation of motion.  Here, $H_{exc}$ is taken in the form given in eq.
(\ref{ham-eff}).  It is easy to verify that, as a consequence, we obtain
\begin{equation}
  i \hbar \frac{\partial a}{\partial t} = \left[ (\tilde{\Omega} - 2 J_0 +
  ) + 2(\chi_1-\chi_2)\, d \left(\frac{\partial u}{\partial
    x} \right)\right]a(x) - \frac{\partial}{\partial x} \left[ (J_0
  +\chi_2 \, d\frac{\partial u}{\partial x} )\left( \frac{\partial a}{\partial
x}\right)
\right] d^2.
  \label{motion-eq}
\end{equation}
We note that the two coupled equations, eqs. (\ref{wave-eq}) and
(\ref{motion-eq}) resemble the Davydov set of equations.  In fact, as $\chi_2
\rightarrow 0$ the result agrees identically with Davydov.

Our next task is to solve these two equations with a specific objective of
finding soliton solutions.

\section{Solving the equations of motion}

In the first step we postulate that
\begin{equation}
  u(x,t) = u(x - vt)
\end{equation}
i.e. that the (real) displacement field is a travelling subsonic wave
($v<v_0$) and
\begin{equation}
  a(x,t) = \exp[- i  E t + i \alpha(x - vt)] a(x - vt)
\end{equation}
which means that the complex exciton field travels with the same velocity $v$
but also has internal oscillations.  The energy of the soliton is not  $E$,
since in our Ansatz, we have other time dependent pieces.  A more complete form
is easy to obtain\cite{davydov}.  As a result, the two equations become
ordinary differential equations in the new coordinate $\xi \equiv x - vt$.
Thus,
\begin{equation}
(v^2 - v_0^2) u^{''} = \frac{1}{M} \left[ 2(\chi_1-\chi_2)\, d a^2 + \chi_2 \,
d^3(a^{'})^2 \right]^{'}
  \label{ode1}
\end{equation}
and
\begin{equation}
  \left[- E + \tilde{\Omega} - 2 J_0 - \frac{v^2}{4 d^2 J_0}+ 2\,
d(\chi_1-\chi_2) u^{'}
\right] a - d^2 \, \left[(J_0 + \chi_2 \, du^{'})a^{'} \right]^{'} = 0
  \label{ode2}
\end{equation}
where the prime denotes differentiation with respect to $\xi$.  Note that as
$\chi_2 \rightarrow 0$, eqs. (\ref{ode1}) and (\ref{ode2}) reduce to the
original system of Davydov equations \cite{davydov}.

To proceed we note that the second term in the square bracket of eq.
(\ref{ode1}) is much smaller than the first one.  This is on the account of the
fact that for large solitons, $a^2 \gg (a^{'})^2$. We can then integrate eq.
(\ref{ode1}) once to yield
\begin{equation}
  u^{'} \cong \frac{2(\chi_1-\chi_2)\, d}{M(v^2-v_0^2)} \, a^2.
\end{equation}
This can then be substituted into eq. (\ref{ode2}) to give
\begin{equation}
  d^2 a^{''} - \frac{\Delta a}{J_0 + \gamma a^2} - \frac{\Gamma a^3}{J_0 +
    \gamma a^2} + \frac{2 d^2  \gamma a}{J_0 + \gamma a^2} (a^{'})^2 = 0.
\end{equation}
where, to simplify notation, we introduced the following symbols
\begin{eqnarray}
  \gamma &\equiv& \frac{2\chi_2(\chi_1-\chi_2)\,d^2}{M(v^2-v_0^2)} \\
\Gamma &\equiv& \frac{4(\chi_1-\chi_2)^2\,d^2}{M(v^2-v_0^2)} +
 \frac{v^2}{4 J_0^2 d^2} \gamma
\end{eqnarray}
and
\begin{eqnarray}
\Delta &\equiv& -E + \tilde{\Omega}  - 2J_0 -  \frac{v^2}{4 J_0 d^2}\, >\,0.
\end{eqnarray}

Eq. (4.6)  has two free parameters, $E$ and $v$ (the velocity of the soliton).
The parameter $E$ is fixed by the requirement that the excitonic wave function
be normalized, ie.
\begin{equation}
\frac{1}{d} \int a^2(z) \; dz = 1.
\end{equation}
Using the transformation
\begin{equation}
  y(a) \equiv \frac{da}{d\xi}
\end{equation}
we obtain from eq. (4.6)
\begin{equation}
y y^{'} + f(a) \, y^2 + h(a) = 0
  \label{bern-eq}
\end{equation}
where
\begin{equation}
  f(a) \equiv \frac{2 \gamma a}{J_0 + \gamma a^2}
\end{equation}
and
\begin{equation}
  h(a) \equiv - \frac{\Delta a + \Gamma a^3}{J_0 +  \gamma a^2}.
\end{equation}
Eq. (\ref{bern-eq}) is a Bernoulli equation which can be completely integrated
and its general solution is
\begin{equation}
y^2 = \exp(-I) (a_1 - 2 \int_{a_0}^{a} h(\tilde{a}) \exp(I) \,d\tilde{a})
  \label{bern-sol}
\end{equation}
where
\begin{equation}
  I \equiv 2 \int_{a_0}^{a} f(\tilde{a}) \, d\tilde{a}.
\end{equation}
In our case
\begin{equation}
  I = 2 \ln (J_0 + \gamma a^2)
\end{equation}
and eq. (\ref{bern-sol}) yields
\begin{equation}
  \left(\frac{da}{d\xi}\right)^2 = \frac{a_1+J_0 \Delta a^2 +
    \frac{1}{2}(\Delta \gamma + J_0 \Gamma) a^4 + \frac{1}{3} \gamma
    \Gamma a^6}{(J_0+\gamma a^2)^2} \equiv H(a).
  \label{final-eq}
\end{equation}
This final equation can be integrated numerically in terms of periodic and
localized solutions.  Note that for $a_1 \neq 0$ only singular or periodic
solutions can be found.  Before we present a complete analysis of the various
solutions we note that in the limit of $\chi_2\rightarrow 0$ we find the
standard elliptic equation \cite{byrd}, i.e.
\begin{equation}
  \left(\frac{da}{d\xi}\right)^2 = \frac{c_0 + \Delta a^2 + \frac{\Gamma}{2}
a^4}{J_0}
\end{equation}
since $\gamma\rightarrow 0$.  Here, elliptic waves can be found in a standard
way, and for $c_0=0$ we find the Davydov soliton as
\begin{equation}
  a(\xi) = \sqrt{- \frac{2 \Delta}{\Gamma}}\; \mbox{sech}\left(\sqrt{
    \frac{\Delta} {J_0}} \xi\right).
\end{equation}

\section{Analysis of solutions}

In solving eq.(\ref{final-eq}), we assumed the values of model parameters
consistent with those used for peptide chains\cite{scott2}. These values are:
\begin{eqnarray}
W & = & 40 \;\mbox{N/m}  \nonumber \\
M & = & 5.7 \cdot 10^{-25} \;\mbox{kg}  \nonumber \\
d & = & 4.5 \cdot 10^{-10} \;\mbox{m} \\
\tilde{\Omega} & \simeq & 0.2 \cdot 10^{-19} \;\mbox{J}  \nonumber \\
J_0 & = & 1.55 \cdot 10^{-22} \;\mbox{J} \label{values} \nonumber
\end{eqnarray}
For illustration purposes, we have chosen to investigate slow solitons ($v/v_0
\simeq 0.02$).  Physical values for $\chi_2$ are on the order of $-1$ pN, and
we will look at values within this range.   Note that $\chi_2$ must be negative
since dipole-dipole interaction decreases with separation.  The value of
$\chi_1$ is still an open question.  Experiments seem to indicate that $\chi_1
\simeq 30-60$ pN.  This would correspond to a compressional soliton.  On the
other hand, theoretical estimates range from $-60 \;\mbox{to} +25 $ pN raising
the possibility of dilatational solitons (for $\chi_1<0$).  Therefore, we will
investigate both positive and negative values for $\chi_1$.

In the case of compressional solitons ($\chi_1>0$),  qualitatively nothing
special happens.  Including the effect of $\chi_2$ simply enhances the
stability of the soliton.  This can be understood as follows.  With $\chi_2 <
0$, we see from eq. (4.7) that $\gamma > 0$ which immediately implies that the
denominator of $H(a)$ given in eq. (4.18) is increased.  This means that in the
compressed chain, the effective $J_0$ is larger.

If $\chi_1 < 0$,  it follows that $\gamma <0$.  In this case, the soliton is
dilated and the dipole-dipole interaction is weakened which can lead to a
destabilization of the soliton itself.

We have examined how the form of the localized solution (soliton) changes with
the value of $\chi_2$.  This is illustrated in Fig. 1, where we see a gradual
sharpening effect at the peak of the soliton solution starting from the
standard Davydov form at $\chi_2 = 0$.   The soliton disappears at a relatively
low value of $\chi_2 \simeq -1.34 pN$, well within the range of physically
acceptable values. This effect has been carefully investigated through analysis
of the numerator and denominator of $H(a)$ appearing in eq. (\ref{final-eq}).
As previously mentioned, $a_1 = 0$ is required for solitary wave solutions to
exist.  Of utmost importance to what follows is the location and reality of the
roots of the numerator in $H(a)$.  Obviously a double root exists at $a=0$, and
the remaining four roots are pairwise symmetrical with respect to sign
reversal.  We shall label the squares of these roots simply as $a^2 = r_1,r_2$.
 With this in place, we note that the singularity in the denominator of $H(a)$
coincides with $r_1 = r_2$ at a point $\chi_2 = \chi_2^{cr}$.  We have
determined that the value of $\chi_2^{cr}$ at  which the solitonic solutions
are destroyed is given by
\begin{equation}
\chi_2^{cr} = \frac{2 J_0 \chi_1}{3 \Delta  - \frac{v^2}{4 J_0^2 d^2} + 2 J_0}.
\end{equation}
Substituting the parameter values from eq. (5.1) into eq. (5.2) we find that
for our system $\chi_2^{cr} = -1.34$ pN which is close to what is physically
expected.  The role of $\chi_2$ in changing the topology of $H(a)$ is clearly
illustrated in Fig. 2.  We show in Fig. 3 the region of stable solitons as
deduced from the relative position of the roots of the numerator of $H(a)$ and
the zero of the denominator.

\section{Conclusion}

This paper has discussed the distance dependence of exciton hopping in the
Davydov model.  Two cases needed to be discussed separately.  In the case of
compressional solitons ($\chi_1>0$), soliton stability was enhanced simply
because in the soliton, the average distance between molecules was decreased,
leading to a higher hopping probability.  If this was the case in peptide
chains, the conclusions reached through thermal stability studies of the
standard Davydov model (Ref. \cite{scott2} and others therein) might be unduly
pessimistic.

The second case of $\chi_1<0$ corresponds to a dilatational soliton.  Here
there exists a critical value of $\chi_2$, beyond which Davydov-like solitons
cease to exist.   This critical value is surprisingly close to physical values
and therefore it is to be expected that thermal fluctuations at physiological
temperatures would destroy the soliton.

By including this important physical effect, we have arrived at a new set of
equations which possess solitons.  These new solutions are clearly relevant to
the physics of poly-peptide chains.

\acknowledgements

This research was supported by an East-West collaboration grant from NATO, the
Alexander von Humboldt Foundation and Deutscher Akademischer
Austauschdienst.  J.A.T. expresses his thanks for the hospitality and kindness
of the staff and faculty members during his stay at the
University of Warsaw.  The authors wish to thank the referee for relevant
criticism which led us to understand the role of the signs of $\chi_1$ and
$\chi_2$.

\begin{figure}
\caption{The plot of soliton profiles for various values of $\chi_2$ (in pN).
The dashed line is the Davydov soliton ($\chi_2 = 0$).  Note how for $\chi_2 =
-1.33$ pN, the soliton becomes sharply peaked.}
\end{figure}

\begin{figure}
\caption{The change in the plot of $H(a)$ for various values of $\chi_2$ (in
pN) corresponding to those used in Figure 1.  The plot with $\chi_2 = 0$ is the
standard Davydov case.}
\end{figure}

\begin{figure}
\caption{The location of squares of the roots, $r_1$ and $r_2$, of the
numerator in $H(a)$ (see eq. (4.18))  and the singularity (dashed curve) in the
denominator of $H(a)$ plotted as a function of $\chi_2$.  Note that
$\chi_2^{cr}$ delineates the boundary of the soliton region.}
\end{figure}

\end{document}